\documentclass[referee]{raa} 
\newcommand{\orcid}[1]{%
    \raisebox{0.7ex}{\scalebox{1}{
        \href{https://orcid.org/#1}{\includegraphics[height=1.5ex]{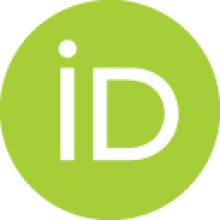}}%
    }}%
}
\usepackage{graphicx,times}             
\usepackage{natbib}
\usepackage{amssymb,amsmath}
\usepackage{CJKutf8}
\usepackage{float}
\bibpunct{(}{)}{;}{a}{}{,}
\usepackage[pagebackref=true]{hyperref}
\usepackage{makecell}
\usepackage{amsmath}

\usepackage{enumitem}

\begin{document}
\begin{CJK}{UTF8}{gbsn}

  \title{\textit{SVOM}/VT:  Flight Model Verification and Pre-launch Testing 
}

   \volnopage{Vol.0 (202x) No.0, 000--000}      
   \setcounter{page}{1}          

   \author{Jian Zhang 
      \inst{1,*}\footnotetext{$*$Corresponding Authors, these authors contributed equally to this work.}
   \and Xue-Wu Fan
      \inst{1}
   \and Gang-Yi Zou
      \inst{1}
   \and Yu-Lei Qiu \orcid{0009-0007-7207-4884}
      \inst{2}
   \and Wei Gao
      \inst{1}
   \and Wei Wang
      \inst{1}
   \and Chen-Jie Wang
      \inst{1} 
   \and Ning Qi
      \inst{1}  
   \and Jin-Song Deng\orcid{000-0001-5646-8583}
      \inst{2,4}      
   \and Li-Jun Dan
      \inst{1}  
   \and Yue Pan
      \inst{1}  
   \and Chao Huang
      \inst{1}  
   \and Yun-Fei Du
      \inst{1} 
   \and Guo-Rui Ren
      \inst{1}        
   \and Zhong-Han Sun
      \inst{1}  
   \and Feng-Tao Wang
      \inst{1}  
   \and Wei Li
      \inst{1}  
   \and Bao-Peng Li
      \inst{1}  
   \and Chao Shen
      \inst{1}  
   \and Peng-Fei Chen
      \inst{1} 
   \and Kun Chen
      \inst{3}
   \and Hui Zhao
      \inst{1} 
   \and Ming Chang
      \inst{1} 
   \and Tao Wang
      \inst{1}       
   \and Li-Pin Xin\orcid{0000-0002-9422-3437}
      \inst{2}
   \and Jian-Yan Wei
      \inst{2}      
   }

   \institute{ Xi’an Institute of Optics and Precision Mechanics, Chinese Academy of Sciences, Xi’an 710119, China; {\it zhjian@opt.ac.cn}\\
        \and
             National Astronomical Observatories, Chinese Academy of Sciences, Beijing 100012, China；\\
        \and
            Innovation Academy for Microsatellites, Chinese Academy of Science, Shanghai 201203, China；\\
        \and
             School of Astronomy and Space Science, University of Chinese Ascademy of Sciences, Beijing 100101, China\\
\vs\no
   {\small Received 202x month day; accepted 202x month day}}

\abstract{ 
This paper presents pre-launch testing and calibration results for the SVOM/VT （Space-based Variable Objects Monitor, Visible Telescope） Flight Model (FM), validating its performance under simulated space conditions through thermal vacuum cycling, energy concentration analysis, stray light suppression, and CCD/electronics calibrations (gain, noise, quantum efficiency). The results confirm full compliance with design requirements: stray light suppression achieves point-source transmittance $<10^{-7}$  at 30° off-axis, thermal control maintains stable CCD temperatures (-75°C for the red channel, -65°C for the blue channel), and detection sensitivity meets the limiting magnitude of 22.50 (SNR \textgreater 3 with 300 seconds exposure). Early in-orbit tests further validate performance, yielding limiting magnitudes of 22.70 (V-band, red) and 22.78 (blue), consistent with pre-launch specifications.
}

   \authorrunning{Jian Zhang }            
   \titlerunning{Flight Model Verification and Pre-launch Testing }  
   \maketitle

%
%
\section{Introduction}           
\label{sect:intro}

\textit{SVOM} is a Sino-French cooperative astronomical project that employs both space-borne instruments and ground-based telescopes to observe Gamma-Ray Burst (GRB) events \citep{2015arXiv151203323C,2016arXiv161006892W,2018SPIE10699E..20G}. Among its four scientific payloads, the Visible Telescope (VT) \citep{2020ApOpt..59.3049F,2020SPIE11443E..0QF} plays a key role in conducting high-sensitivity, two-channel (blue and red band) optical follow-up observations of gamma-ray bursts (GRBs). It also identifies high-redshift candidates to facilitate further near-infrared observations using large ground-based telescopes.

In this section, we briefly introduce the composition of VT, its operation modes and key specifications.
\subsection{Composition of the VT }

\begin{figure}
    \centering
    \includegraphics[width=1\linewidth]{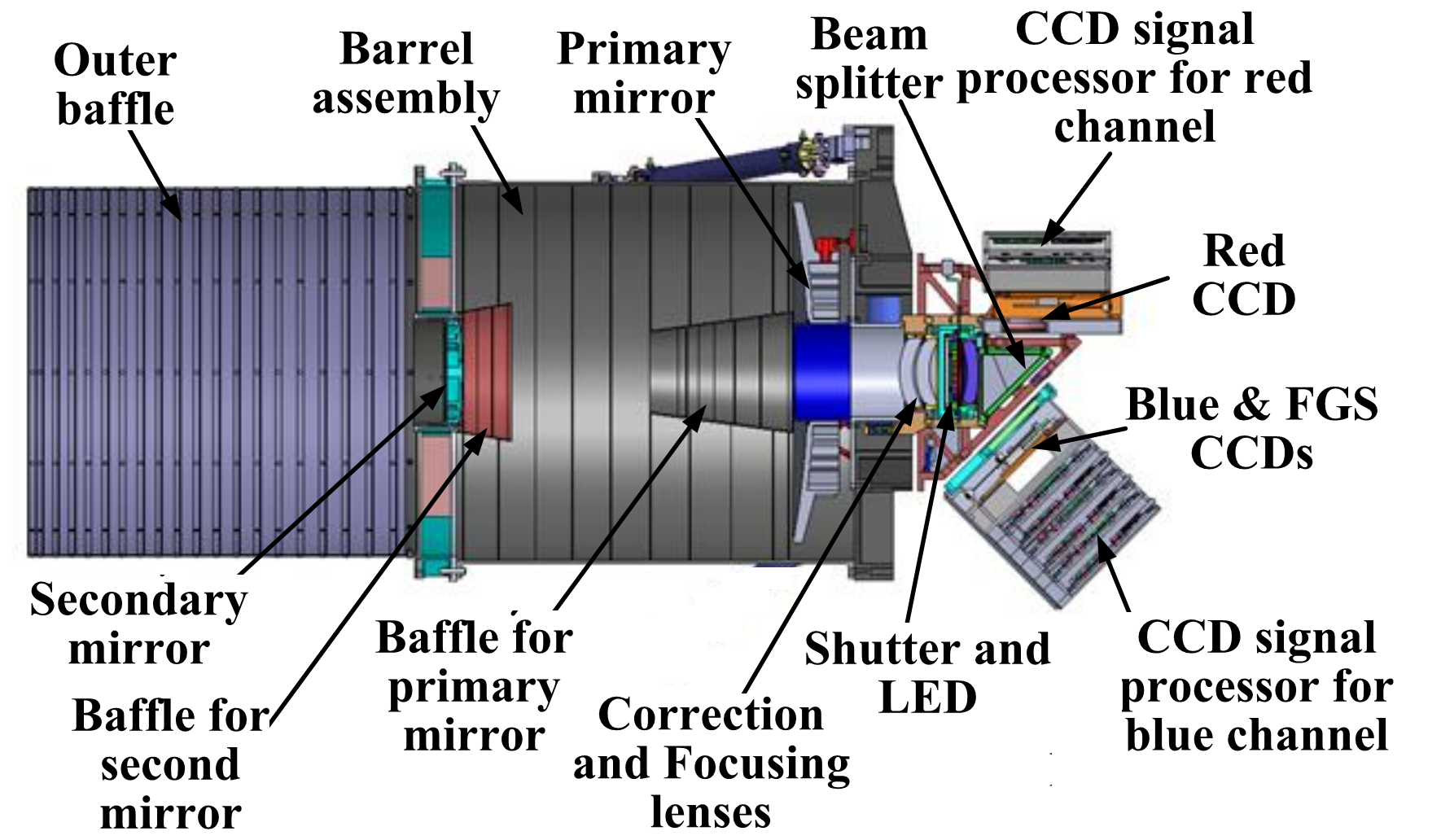}
    \caption{Composition of the VT main body.}
    \label{fig1}
\end{figure}

Figure \ref{fig1} shows a schematic diagram of the VT, outlining its key components—from the optical system to the CCD cooling system.
 
\textbf {Optical System} The VT employs a 44 cm aperture Ritchey-Chrétien (RC) optical system (f/8 focal ratio). Its lightweight silicon carbide (SiC) primary mirror, coated with high-reflectivity silver, achieves a surface accuracy less than 1/50 $\lambda$. The secondary mirror is optimized to support the wide field of view (FOV) required for mounting two Fine Guidance Sensors (FGS) alongside the main CCD in the blue channel's focal plane.

\textbf{Dual-Channel Imaging} A dichroic beam splitter divides incident light into red (400–650 nm) and blue (650–1000 nm) channels, directing each to dedicated CCDs. The beam splitter exceeds 95\% transmittance and reflectivity efficiency in both bands, ensuring precise co-alignment and maximizing throughput for rapidly evolving transients like gamma-ray bursts (GRBs).

\textbf{Detectors} Both channels provide a $26^{\prime}$ × $26^{\prime}$ FOV at $0.77^{\prime\prime}/pixel$, utilizing e2v 42-80 2048 × 2048 CCDs(13.5 µm pixels) optimized for their respective bands:
\begin{itemize}
    \item Red Channel: Deep-depletion, back-illuminated CCD (QE $\sim 60\%$  at 900 nm), operating in Non-Inverted Mode (NIMO).
    \item Blue Channel: Thinned, back-illuminated CCD (QE $\sim 90\%$ at 550 nm), operating in Advanced Inverted Mode (AIMO).
\end{itemize}
	
\textbf {Calibration System}
The integrated calibration unit acquires bias, dark, and flat-field frames using a motorized shutter coated with diffusing material, which acts as a flat-field source when illuminated by LEDs. For the blue channel, LEDs emit wavelengths of 470nm, 527nm, and 640nm, while the red channel uses 670nm, 740nm, and 870nm. When the shutter is closed, the system captures bias frames (zero-second exposures) and dark frames (LEDs off), and when the shutter is illuminated, multi-wavelength flat-field frames (LEDs on) are obtained.

\textbf {Stray Light Suppression} The primary and secondary mirror baffles prevent skylight from directly illuminating the detectors without first reflecting off the mirrors. The outer baffle and tube block off-axis light (beyond 24°) from reaching the primary mirror. To further suppress stray light, all inner surfaces of the baffles and tube are coated with anti-reflective materials and fitted with light-trapping vanes. When light from bright off-axis sources enters the telescope, it undergoes multiple reflections within the outer baffle and tube before reaching the primary mirror. This trapped light is further attenuated by the anti-reflective coatings and baffle vanes on the inner surfaces of the primary and secondary mirror baffles, minimizing stray light contamination.

\textbf {Focusing and Aberration Correction}
To correct the residual aberration of the RC configuration, the VT employs three lenses positioned near the primary mirror (PM). The first two lenses are motor-driven and move along the optical axis to adjust focus.

\begin{figure}[h]
    \centering
    \includegraphics[width=0.9\linewidth]{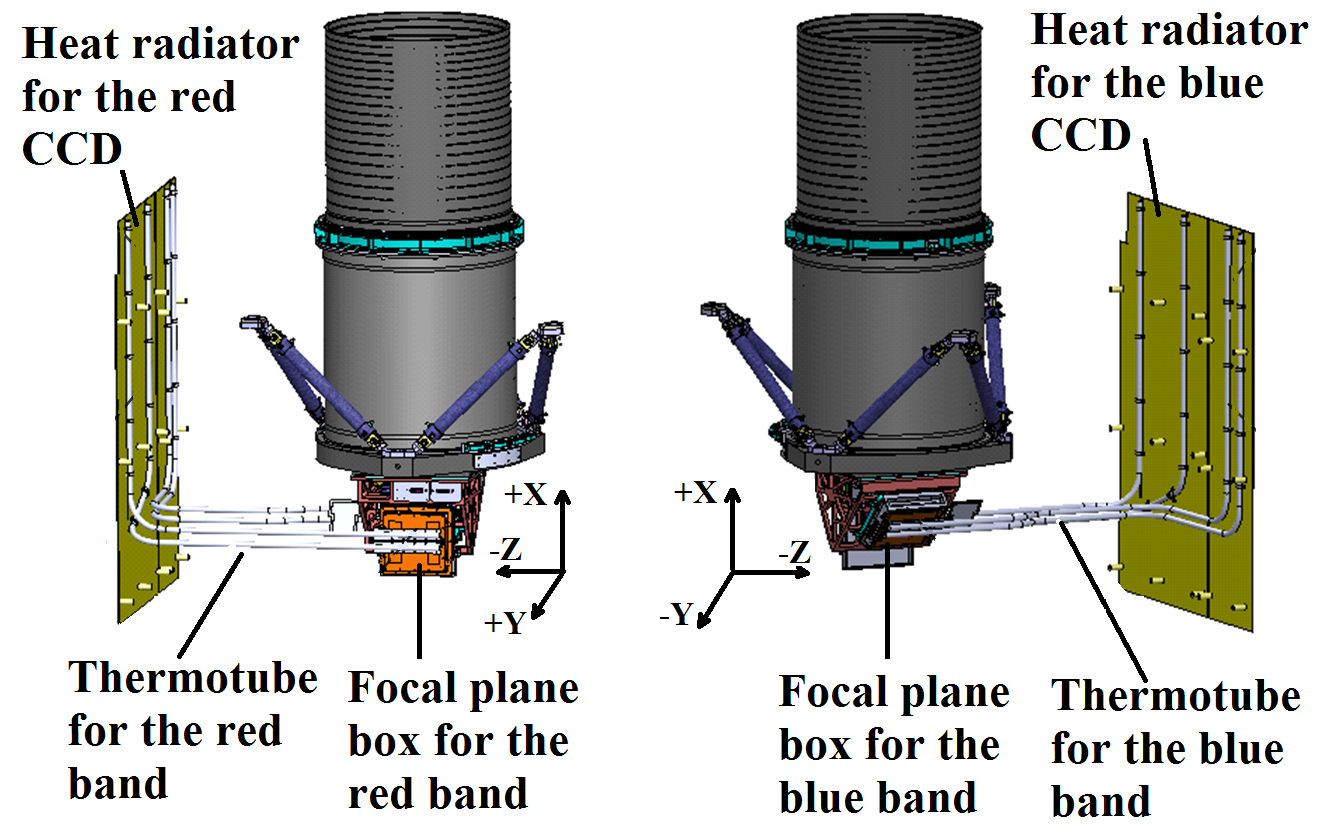}
    \caption{Heat dissipation path of the CCD. Left: red band, Right: blue band.}
    \label{fig2}
    \vspace{-10pt}
\end{figure}

\textbf {The Cooling system for CCDs} To decrease the dark current, the working temperatures of CCDs must be lowered to an appropriate level. Due to differences in operation modes between the two CCDs, the dark current in the red-band CCD (operated in Non-Inverted Mode, NIMO) is significantly higher than that of the blue-band CCD (operated in Advanced Inverted Mode, AIMO) at the same temperature. To mitigate this, the operating temperature of the red-band CCD (-75°C) is set lower than that of the blue-band CCD (-65°C). Each CCD uses a two-stage Thermo-Electric Cooler (TEC) which connects to a dedicated heat radiator with heat pipes for cooling. A schematic diagram of the CCD temperature control design is shown in Figure  \ref{fig2}.

\subsection{Operation Modes in Orbit}

\begin{figure}[h]
    \centering
    \includegraphics[width=0.9\linewidth]{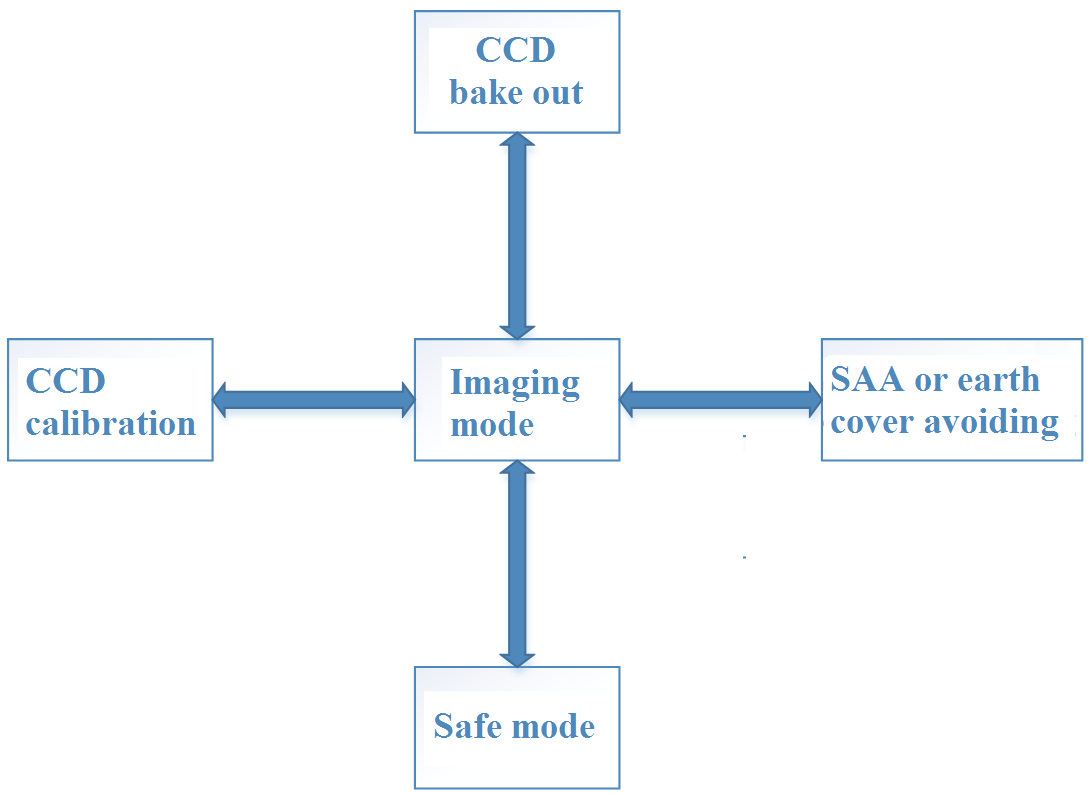}
    \caption{VT operation modes in orbit. The VT is within the imaging mode at most of the time, expect for some reason it could change its woking statue to another mode.}
    \label{fig3}
\end{figure}

Below is a brief overview of VT’s primary operation modes (see Fig. \ref{fig3})

\textbf{Imaging Mode} 
The default mode for observations, configurable via parameters such as exposure time, readout mode, readout speed, and window size.

\textbf {SAA and Earth Occultation Modes}
Automatically deactivates CCD sensors when the satellite enters the South Atlantic Anomaly (SAA) region or Earth’s occultation to protect hardware.

\textbf{Calibration Mode} 
Performs periodic in-flight calibrations to correct radiation-induced hot pixels and maintain data quality.

\textbf{Bake-out Mode} 
Mitigates radiation damage by heating the detector to 25°C (CCD annealing) when required.

\textbf {Safe Mode}
Activated during anomalies, this mode powers down non-critical systems while maintaining thermal control. Optical components are stabilized at safe temperatures, and CCDs remain passively cooled.

\subsection{Key Specifications}

\begin{table}[h]
    \vspace{-10pt}
    \centering
    \caption{VT Key Specifications}    
    \begin{tabular}{ccc}
      \hline\noalign{\smallskip}
        No &Item  & Specification\\
      \hline\noalign{\smallskip}
        1 & Telescope aperture & 440 mm \\
        2 & Focal length & 3600mm \\
        3 & Field of view & $26^{\prime}$×$26^{\prime}$ \\
        4 & Filter band-pass & 400nm-650nm, 650nm-1000nm \\
        5 & \makecell[c]{Encircled energy \\for red}  &  \makecell[c]{ radius of EE70 \\within 1.0 arcsecond}\\
        6 & \makecell[c]{Encircled energy \\for blue} & \makecell[c]{The radius of EE80 \\within 1.5 arcsecond}\\
        7 & \makecell[c]{Stray light suppression} & \makecell[c]{\(< \)1/3 sky background \\when the moon is\\ \(>\) 30° off optical axis}\\
        8 & \makecell[c]{Red CCD working \\temperature} & -75°C±2°C \\
        9 & \makecell[c]{Blue CCD working \\temperature} & -65°C±2°C \\
        10& Readout noise &  \makecell[c]{\(<\)6e$^{-}$/pix @100KHz} \\

      \hline\noalign{\smallskip}
    \end{tabular}
    \label{tab:1}
    \vspace{-10pt}    
\end{table}

Table \ref{tab:1} summarizes the key specifications of the VT, as tested and calibrated on the FM. The pre-launch testing procedures, calibration methods, and their results are detailed in the following section. 

\section{The pre-launch experiments, tests and calibrations}
\label{sect:Obs}
\subsection{The Thermal Vacuum Experiment}

\subsubsection{Thermal Environment Analyses of the VT in Orbit}
 
The thermal vacuum experiment aims to validate the VT's operational adaptability to varying thermal conditions in space.

\begin{figure}
    \centering
    \includegraphics[width=1\linewidth]{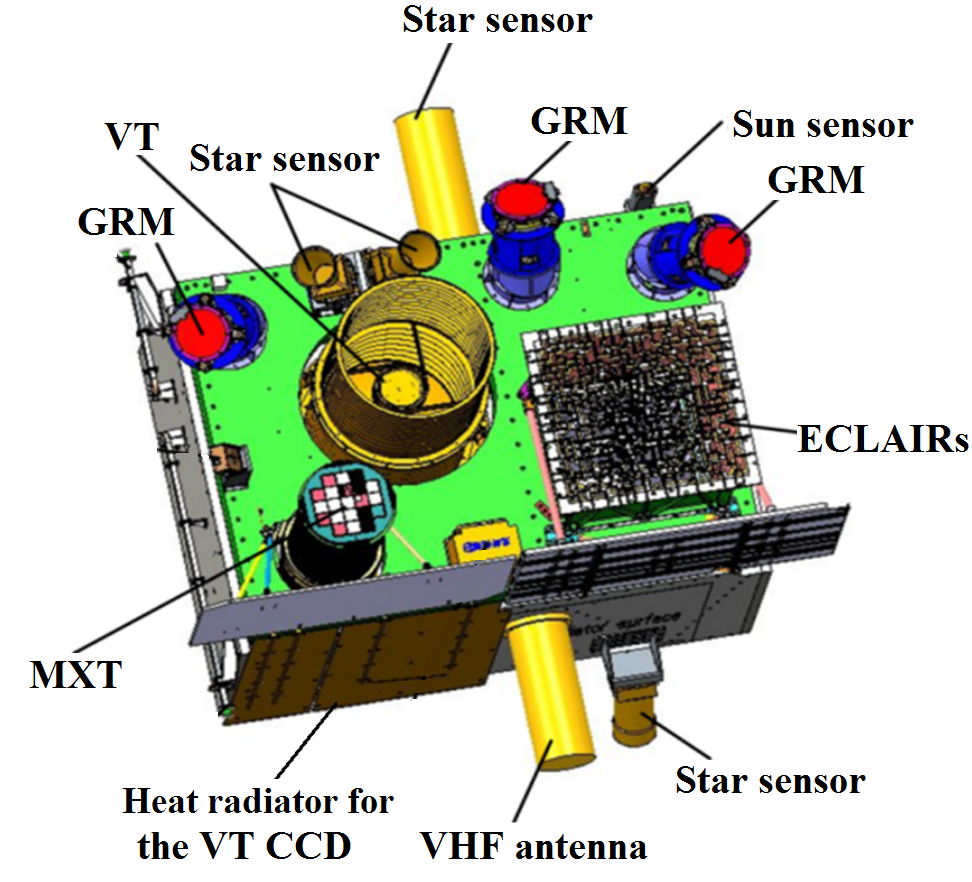}
    \caption{The external view of the satellite: the most part of the VT is in the satellite, but the outer baffle and the heat sink plate is outside}
    \label{fig30}
\end{figure}

After assembly, most of the VT’s opto-mechanical structure (see Fig. \ref{fig30}) is housed within the thermally stable payload bay. However, the outer baffle of the payload module is exposed to variable heat flux—particularly when pointing toward cold space or entering Earth’s occultation region, where the aperture faces the warmer Earth. During attitude maneuvers—particularly when crossing the day-night boundary—the external heat flux varies abruptly, necessitating strict thermal stability of the entire opto-mechanical assembly.

Meanwhile, the heat radiators may intermittently face the warmer Earth, causing thermal load fluctuations. The TECs must compensate for these variations to stabilize the detector's temperature.

To validate the VT's thermal control capability in space under the complex conditions mentioned above, we designed five test scenarios:
\begin{enumerate}[label=(\arabic*)]
\item Nominal thermal state: The VT remains oriented toward deep space, outside Earth's occultation.
\item Minimum heat flux exposure: The radiator receives the lowest achievable heat flux under steady-state conditions.
\item Maximum heat flux exposure: The radiator operates under peak heat flux, approaching its saturation limit.
\item Heat flux surge during maneuvers: During attitude adjustments, the VT pupil experiences a rapid surge, transitioning from minimum to maximum flux levels.
\item Heat flux drop during maneuvers: Post-maneuver, the VT pupil undergoes an abrupt decline, falling from peak to near-minimal flux.
\end{enumerate}

\subsubsection{Thermal Vacuum Experiment Setup}

\begin{figure*}
    \centering
    \includegraphics[width=1\linewidth]{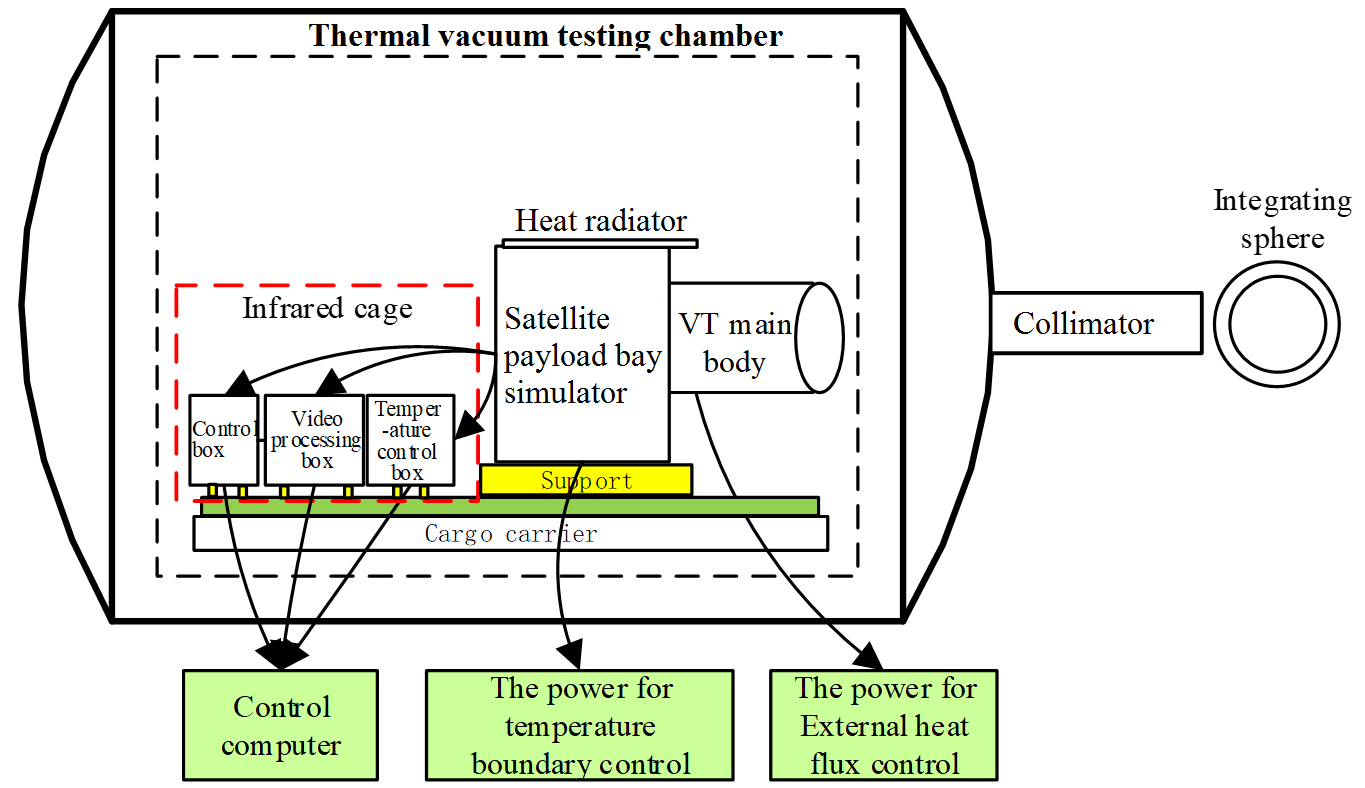}
    \caption{Setup of the VT thermal vacuum experiment. }
    \label{fig4}
\end{figure*}

The setup of the thermal vacuum experiment, which validated the VT's adaptability to the varying thermal environments of space, is shown in Figure  \ref{fig4}. In this experiment, the VT main body was assembled with the satellite payload module simulator, except for the outer baffle and part of the barrel assembly, which extended out to face a 30-meter focal length collimator. The satellite module simulator and other electronics boxes were placed inside the thermal vacuum testing chamber, while other testing equipments remained outside. Cables passing through the tank facilitated communication between the interior and exterior. An infrared cage heater and thin-film heaters attached to the outer surface of the payload simulator and the outer baffle created a simulated in-orbit thermal vacuum environment. A simulated star point at infinite distance was created using the collimator, an integrating sphere, and a star point with a diameter of 0.02 mm.

\subsubsection{Thermal Vacuum Experiment Results}

During the thermal vacuum test, all five thermal environment scenarios were evaluated. The temperatures of the VT FM’s key components remained stable, with fluctuations confined to a narrow range that met and exceeded the design requirements, as detailed in Table \ref{tab:2}.

\begin{table}[h]
    \centering
    \caption{The Key Components of the FM of VT Temperature Status in Thermal Vacuum Experiment }
    \begin{tabular}{ccc}
        \hline\noalign{\smallskip}
         Item&  Requirement (℃)& Experiment result (℃)\\
        \hline\noalign{\smallskip}
         Primary mirror&  20±2& 20.3 $\sim$ 20.6\\
         Secondary mirror&  20±2& 19.9 $\sim$ 20.3\\
         Barrel assembly&  20±2& 19.9 $\sim$ 20.1\\
         Red band CCD&  -75±2& -75.2 $\sim$ -74.8\\
         Blue band CCD&  -65±2& -65.2 $\sim$ -64.8\\
      \hline\noalign{\smallskip}
    \end{tabular}
\label{tab:2}    
\end{table}

\subsection{Energy Concentration Test}

\begin{figure}[h]
    \centering
    \includegraphics[width=1\linewidth]{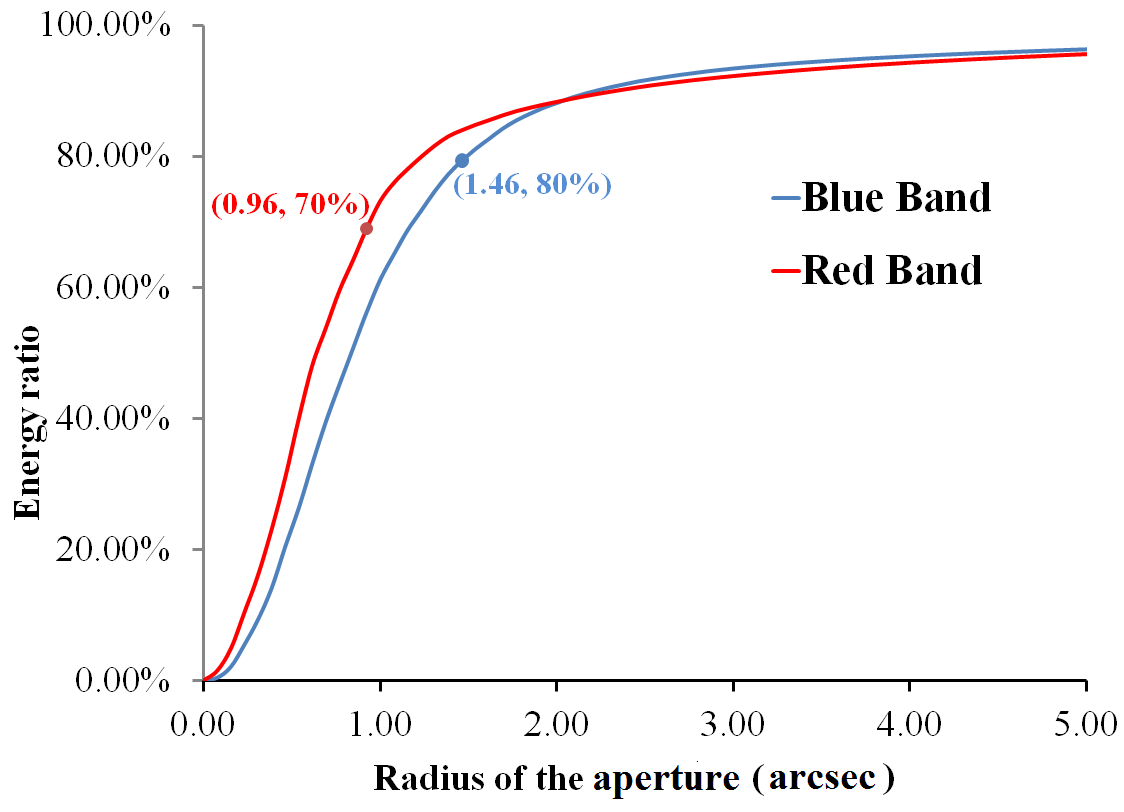}
    \caption{Energy growth curves of the red band and the blue band}
    \label{fig7}
\end{figure}

An energy concentration test was conducted during thermal vacuum testing to evaluate the point spread function (PSF), a critical parameter for imaging quality, using encircled energy (EE) as the metric. Figure \ref{fig7} illustrates the growth of EE in the red and blue bands as a function of aperture size. The results demonstrate an EE70 radius (70\% encircled energy) of $0.96^{\prime\prime}$ for the red band and an EE80 radius (80\% encircled energy) of $1.46^{\prime\prime}$  for the blue band. These values meet the requirements specified in Table \ref{tab:1}  ($1.0^{\prime\prime}$ and $1.5^{\prime\prime}$ , respectively). 

\subsection{Point Source Transmittance Test}

According to design requirements, when the full Moon is 30° off the VT's optical axis, the stray light brightness on the focal plane should not exceed one-third of the skylight background. This is equivalent to a point source transmittance (PST) of less than $1.32 \times 10^{-6}$ \citep{2020SPIE11443E..0QF}.

A PST (Point Spread Function) test was performed on the FM after opto-mechanical assembly. The test methodology followed the approaches described by \citet{2008SPIE.7069E..0OF} and \citet{2010SPIE.7794E..0WG}, using a 660 nm light source in a double-column tank. Identical testing conditions to those applied to the Qualification Model (QM) were maintained to ensure consistency.

\begin{figure}[h]
    \centering
    \includegraphics[width=1\linewidth]{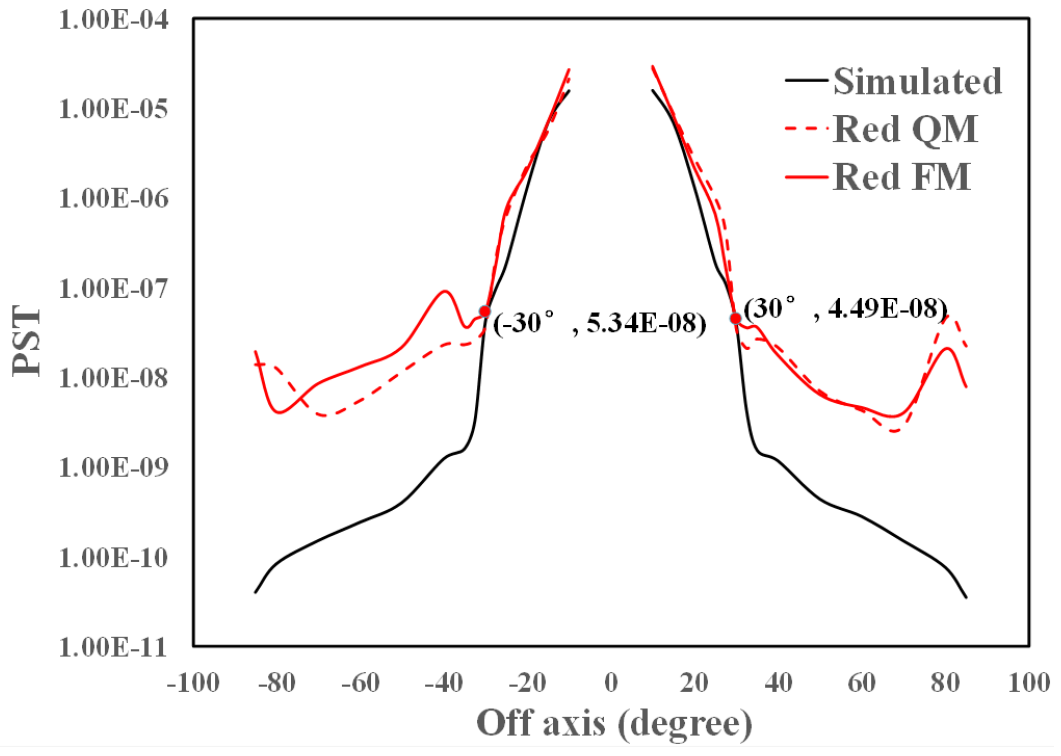}
	\caption{\label{Fig8}{Red-band PST curves from the FM and QM tests and design simulation} }
\end{figure}

\begin{figure}[h]
    \centering
    \includegraphics[width=1\linewidth]{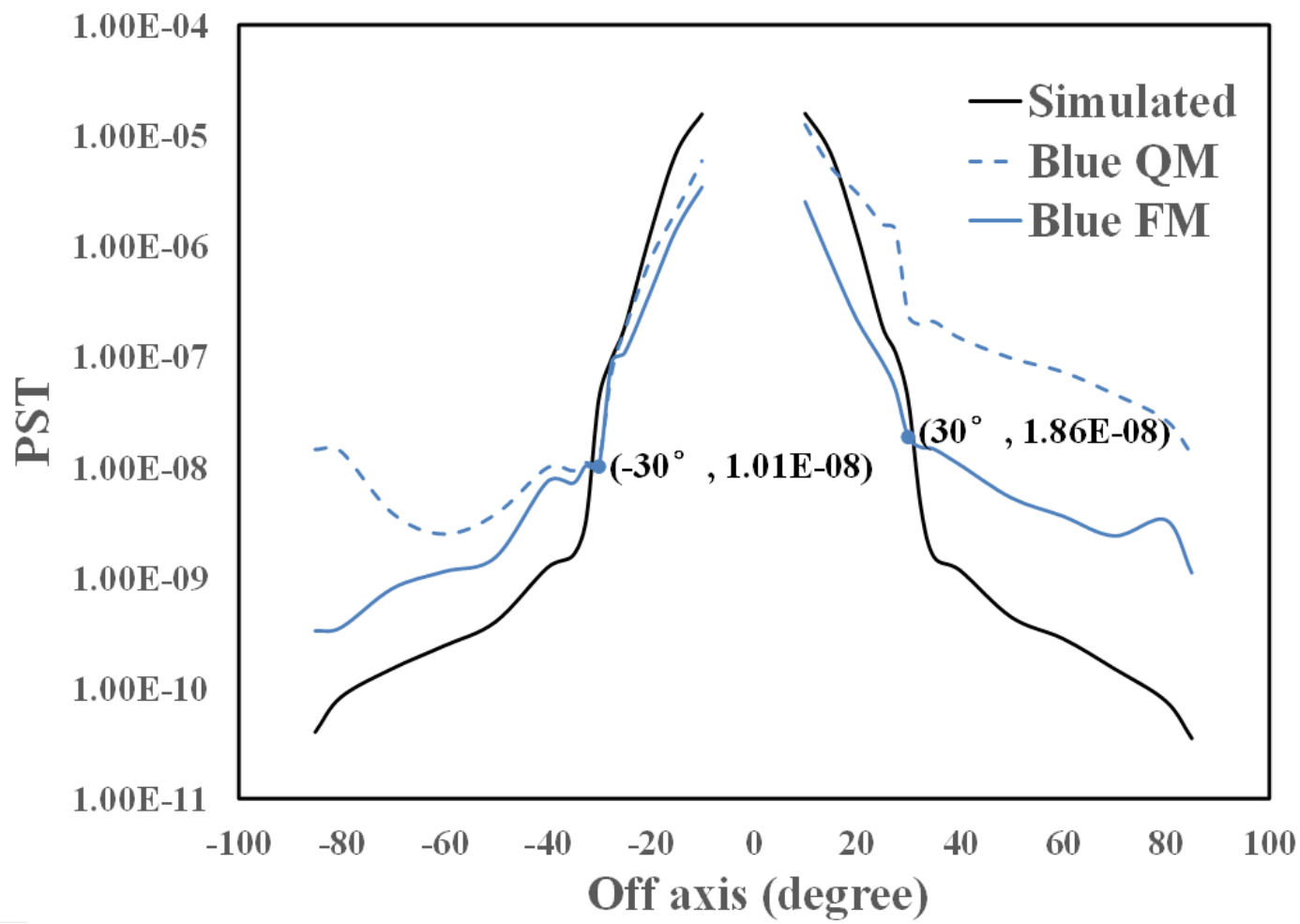}
	\caption{\label{Fig9}{Blue-band PST curves from the FM and QM tests and design simulation} }
\end{figure}

The results, compared with QM and design simulations, are presented in Figure \ref{Fig8} (red band) and Figure \ref{Fig9} (blue band). As stray light intensity decreases, ambient background light contributes to higher measurement uncertainty. Notably, when the point source transmittance (PST) drops below $1.00 \times 10^{-7}$, this uncertainty can increase by up to 10 times the nominal value. The FM stray light suppression performance closely matches that of the QM, remaining within the anticipated margin of error. At 30 degrees off-axis, the FM’s PST consistently stayed below $10^{-7}$ in both bands, fully complying with all design requirements. 

\subsection{Calibrations on CCDs and their Electronics}

\begin{figure}[h]
    \centering
    \includegraphics[width=1\linewidth]{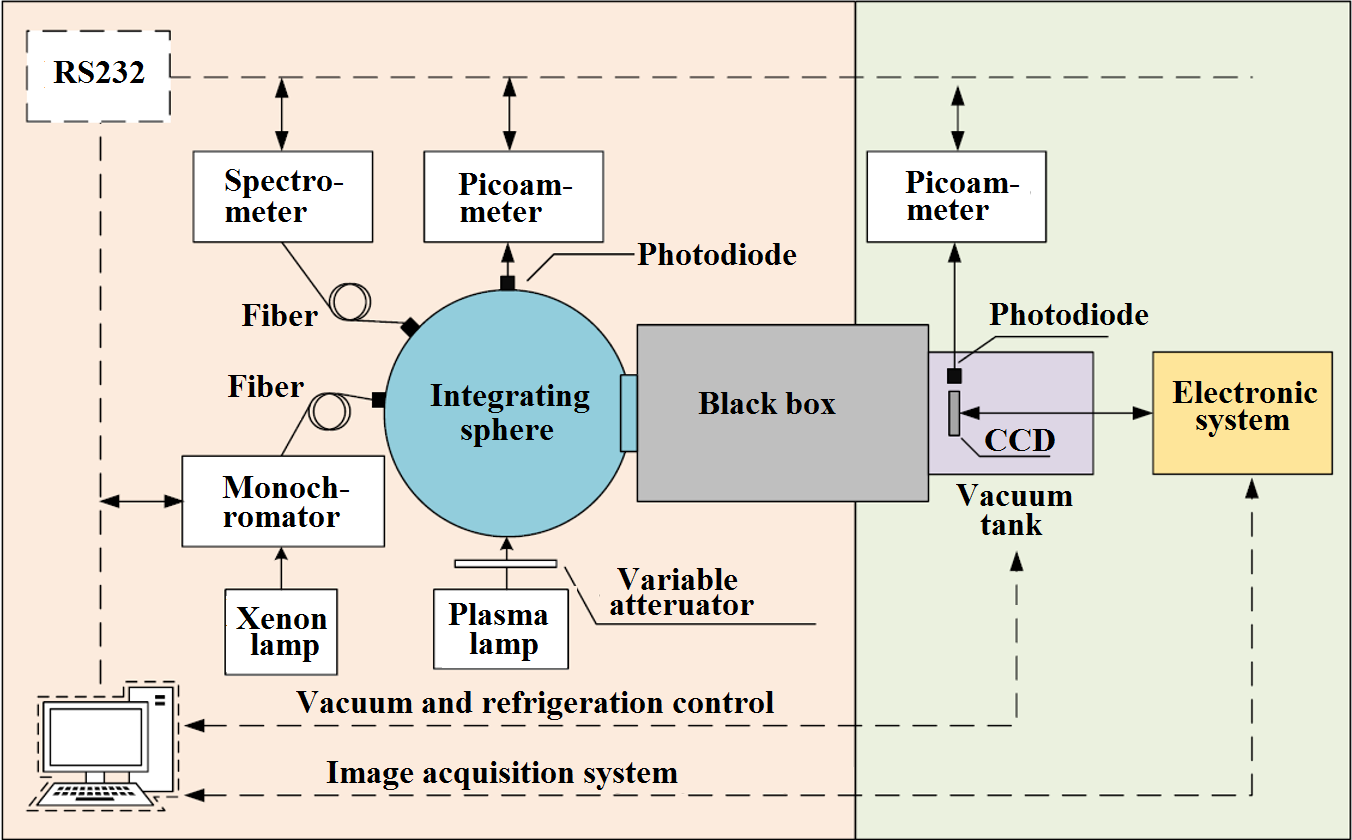}
    \caption{Setup of the CCD calibration platform}
    \label{fig10}
\end{figure}

CCD calibrations were conducted after circuit debugging, which included gain, quantum efficiency (QE), readout noise, and dark current 
The setup of the calibration platform is shown in Figure \ref{fig10}.

\begin{figure}[h]
    \centering
    \includegraphics[width=1\linewidth]{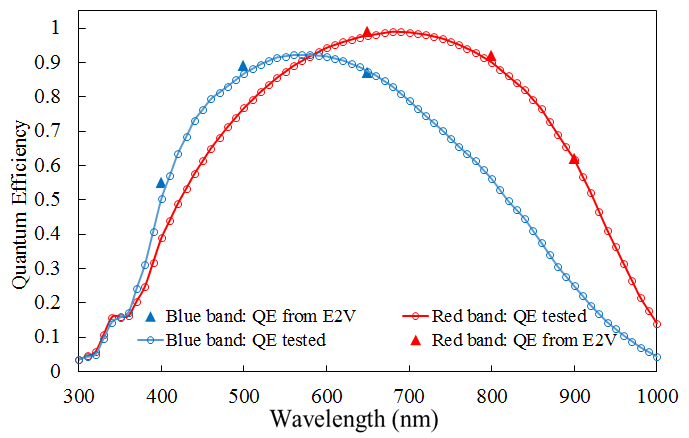}
    \caption{CCD QE tested results and from manufacturer E2V in the red band and blue band. The resule in calibration is consistent with the manufacturer's results.}
    \label{fig11}
\end{figure}

During calibration, the CCDs were placed in a vacuum tank with sensor temperatures set to in-orbit conditions: -75°C for the red band and -65°C for the blue band. Figure \ref{fig11} presents the QE results for both bands, while Table \ref{tab:3} summarizes the measured gains, readout noise, and dark currents at a readout rate of 100 kHz. The tests confirmed that the FM performance aligns with both the QM \citet{2021Electronics10.931,2021Photo...8..132P,2023SPIE12965E..0BP,2022PASP..134c7001P} and the design requirements. 

\begin{table}[h]
    \centering
\caption{CCD electronics calibration results}
\label{tab:3}
    \begin{tabular}{cccc}
        \hline\noalign{\smallskip}
         Spectral band&  Red &  Blue\\
        \hline\noalign{\smallskip}
         Gain (e/DN)&  2.293 ± 0.003&  1.949 ± 0.002\\
        \hline\noalign{\smallskip}
         read-out noise (e/pixel)&  5.62 ± 0.02&  5.10 ± 0.01\\
        \hline\noalign{\smallskip}
        Dark current (e/s/pixel)& 0.068 ± 0.003& 0.003 ± 0.001\\
        \hline\noalign{\smallskip}
    \end{tabular}
\end{table}

\subsection{System Optical Efficiency Calibration}

Optical Efficiency (OE) characterizes the ability of the whole telescope system to effectively collect incoming photons and convert them into measured photoelectrons. OE is affected by the transmittance/reflectance of optical components and the QE of CCDs. The result will be used to estimate detection capabilities of the telescope under various assumptions for the target and background characteristics. 

The OE calibrations were performed using two methods: 

The first method involved separately measuring the optical reflectance/transmittance of each component after optical coating and then synthesizing the OE by multiplying these values with the independently measured QE.

The second method directly measures the assembled system to obtain the directed OE. This process can be divided into two individual steps.

\begin{itemize}
    \item \textbf{ Step 1: Light Source Calibration} \\
     The test setup is illustrated in Figure \ref{fig12}. The light from a xenon lamp, filtered through a monochromator, enters an integrating sphere and exits via a star point positioned at the focal point of the collimator. The collimator converts this light into a parallel beam, which serves as the input source for telescope calibration.

The goal of this test is to calibrate the collimator’s output light flux. Direct measurement of the collimated light intensity is impractical due to its extremely low signal level. To address this, we employ an off-axis parabolic mirror to integrate the collimated beam onto a photo-detector (picoammeter). Simultaneously, another high-sensitivity photo-detector inside the integrating sphere measures the internal light intensity. By correlating the readings from both detectors, we establish the relationship between the total collimator output intensity and the integrating sphere’s internal intensity. In this way, the input source for the telescope has been calibrated.

\item \textbf{ Step 2: Telescope Calibration} \\
For this calibration, we replace the reflective mirror used in the previous test with the VT (Fig. \ref{fig12}). The VT focuses the collimated light into a circular spot at its focal planes, received by the two CCD detectors. The relative OE is calculated as the ratio of the total flux on the focus plane to the flux entering the pupil. To determine the absolute OE, we scale this result to the VT’s full aperture using a correction factor derived from the ratio of the VT’s pupil surface area to that of the collimator.
\end{itemize}
\begin{figure*}
    \centering
    \includegraphics[width=0.8\linewidth]{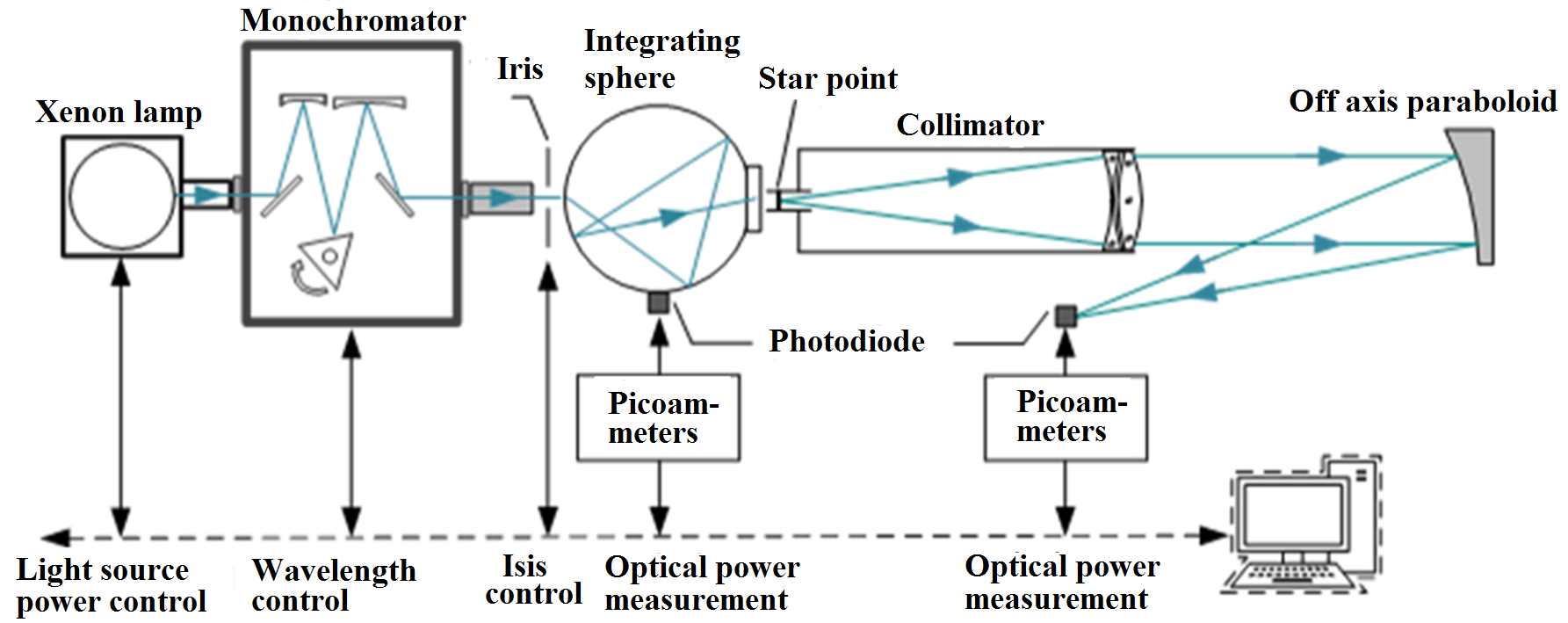}
    \caption{The calibration of light source }
    \label{fig:placeholder}
\end{figure*}

\begin{figure*}[h]
    \centering
    \includegraphics[width=0.8\linewidth]{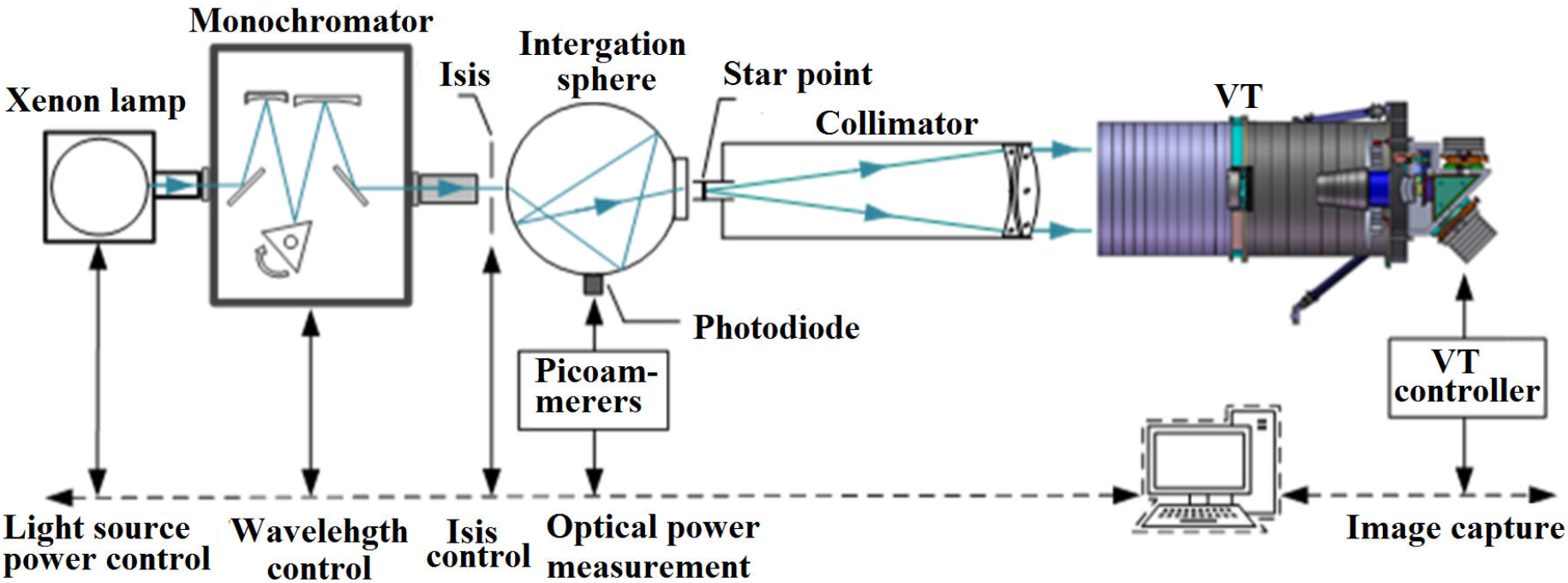}
    \caption{Setup for direct calibration of optical efficiency }
    \label{fig12}
\end{figure*}

\begin{figure}
    \centering
    \includegraphics[width=1\linewidth]{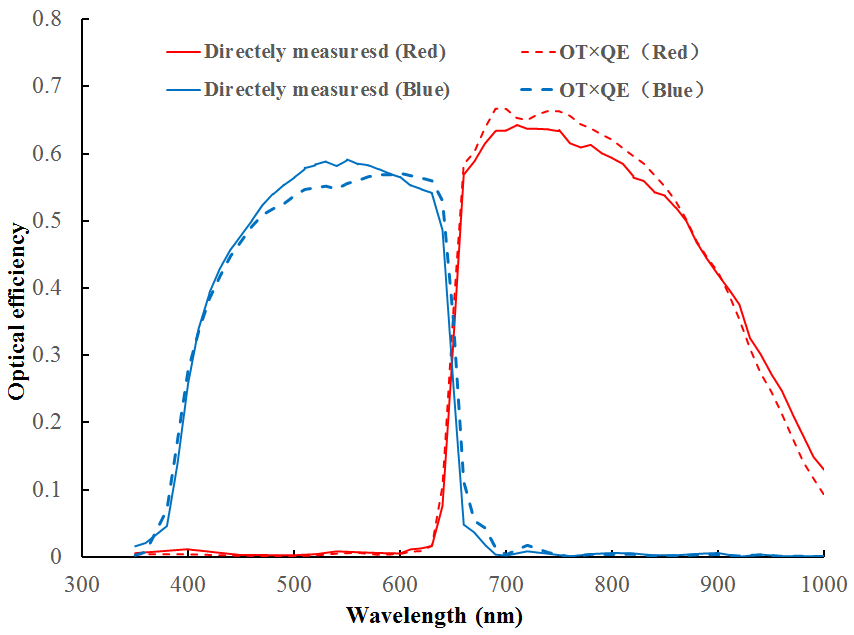}
    \caption{Optical efficiency calibration results in the red band and blue band}
    \label{fig13}
\end{figure}

The results from the two methods are consistent and are shown in Figure \ref{fig13} for both bands. The maximum relative deviation
between the two result is less than 5\% for both bands.

\section{Estimation of detection capabilities}
\label{sect:data}

According to the design requirements, the signal-to-noise ratio (SNR) for a 300-second observation of GRB events at magnitude 22.5 should not be less than 3. In practice, to mitigate the effects of satellite jitter, three consecutive 100-second exposures taken by the VT in orbit are co-added to approximate an equivalent single 300-second observation.

Since directly validating the VT  detection on the ground for stars of magnitude 22.5 is challenging, its detection capabilities were derived by synthesizing the calibration results with the spectral characteristics of typical GRB events and the skylight background. The SNR was calculated using Equation \ref{eq1}.

\begin{equation}\label{eq1}
\text{SNR} = \frac{\sqrt{N}C R_{*} t}{\sqrt{ C R_{*} t + n_{\text {p}} \left( R_{\text{s}} t + R_{\text{d}} t + (R_{\text{r}} + G/2)^2 \right) }}
\end{equation}

where:
\begin{itemize}
    \item \(R_{\text {r}} \) is the readout noise per pixel.
    \item \textit{G} is the electronics gain.
    \item \(R_{\text {d}} \) is the dark current rate per pixel.
    \item \textit{C} is Energy concentration factor (80\% blue, 70\% red).
    \item \(n_{\text {p}} \) is the number of  pixels for a GRB point-like source.
    \item \textit{N=3} is the number of exposures.
    \item \textit{t=100} is the exposure time for a single exposure.
    \item \(R_{\text {*}} \) is the count rate of GRB photoelectrons.
    \item \(R_{\text {s}} \) is the count rate of  photoelectrons of a single pixel due to sky background and stray light.    
\end{itemize}
Due to satellite jitter (which can reach up to $1.6^{\prime\prime}$ in the worst-case scenario), the in-orbit PSF of a point-like source will be broader than its laboratory-measured size.
Assuming Gaussian distributions for both photon distribution of the point-like sources and satellite stability—and incorporating laboratory test results—the maximum in-orbit radii are calculated as  $1.75^{\prime\prime}$  for EE80 (blue band) and  $1.27^{\prime\prime}$ for EE70 (red band). Given the VT’s pixel angular resolution of  $0.77^{\prime\prime}$ , the corresponding number of pixels $n_{p}$ required to enclose these radii are 16.18 (EE80, blue) and 8.54 (EE70, red).

We assume that the GRB spectrum follows a power-law distribution (Equation \ref{eq2}), adopting a typical value of ($\alpha = 1$) for simplicity.  The count rate of photoelectrons $R_{\text {*}}$ generated by the VT for GRBs of varying brightness levels are calculated by combining the GRB spectrum, telescope aperture, and our measured OE values.

\begin{equation}\label{eq2}
    F_{\lambda} \mathrm{d} \lambda\propto\lambda^{\alpha-2} \mathrm{~d} \lambda
\end{equation}

The background count rates are estimated at 1.46 photon/pixel/s for the red band and 1.21 photon/pixel/s for the blue band, assuming that the background is dominated by the zodiacal light. To account for stray light from the full Moon (assuming $30^\circ$ off-axis), the sky count rate $R_{\text {s}}$ is further multiplied by a factor of 1.33.

\begin{figure}
    \centering
    \includegraphics[width=1\linewidth]{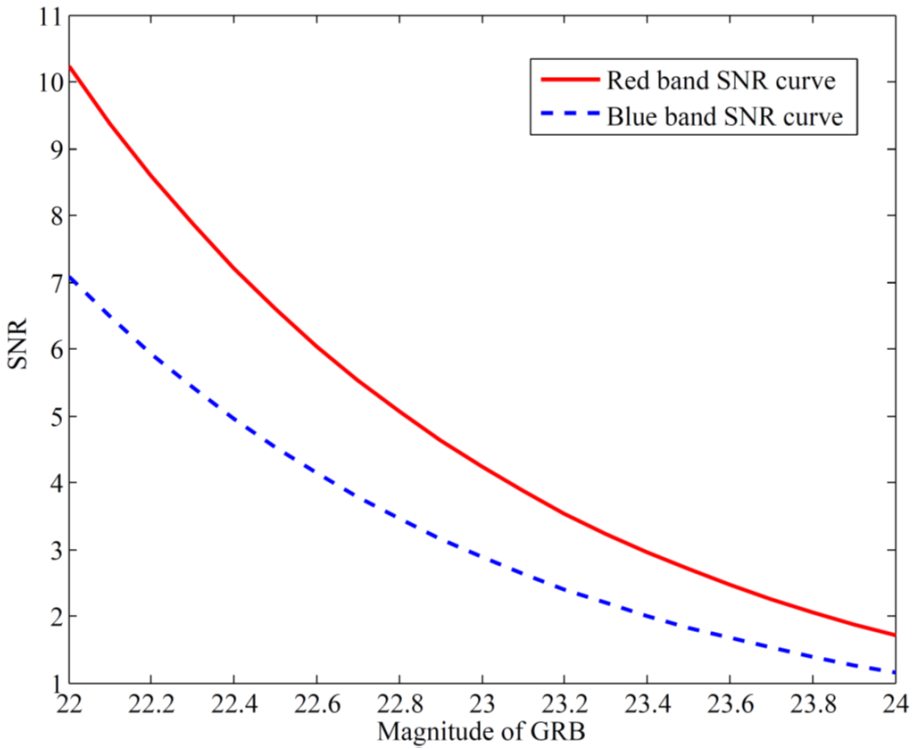}
    \caption{The VT SNR curves for GRBs of varying brightness levels. At magnitude of 22.5, the SNR is 6.6 for the red and 4.5 for the blue bands.}
    \label{fig14}
\end{figure}

The simulated VT SNR curves, for GRBs of varying brightness levels, are shown in Figure \ref{fig14}. At magnitude of 22.5, the SNRs for the red and blue bands are 6.6 and 4.5, respectively, well meeting the design requirements.

\section{ Conclusions}
\label{sect:conclusion}

The VT is a critical satellite payload for conducting follow-up optical observations of \textit{SVOM} GRBs. To validate its in-orbit performance prior to launch, extensive testing—including optical efficiency measurements, thermal tests, imaging tests, straylight suppression assessments, and detection capability evaluations—was conducted under simulated space conditions. All test results confirm that the VT flight model meets the required design specifications.

During the commissioning phase and subsequent operational observations after \textit{SVOM}'s launch, the VT instrument demonstrated performance exceeding expectations \citep{vt_overview_qiu+etal+2026,Yao+etal+2026}. The point spread functions (PSFs) in both channels remain highly stable under complex thermal conditions, and the focus exhibits long-term stability, requiring only one minor adjustment since launch in June 2024. Contamination has been effectively controlled, with minimal impact on performance. CCD temperatures are tightly stabilized, and straylight suppression meets specifications, as confirmed by off-axis tests under full Moon illumination. The satellite’s exceptional stability \citep{lidong+etal+2026} further minimizes platform contributions to the PSF. The measured (3σ) limiting magnitudes are 22.70 (V-band equivalent) in the red channel and 22.78 in the blue channel for a 300-second exposure.

VT has also demonstrated outstanding GRB detection efficiency for both \textit{SVOM} and external triggers. Notably, it contributed to the selection of high-redshift candidates, including GRB 250314A at z = 7.3, by providing deep upper limits \citep{2025arXiv250718783C}. These results collectively confirm that VT not only meets but surpasses its design requirements.

\begin{acknowledgements}
The \textit{SVOM} is a joint Chinese-French mission led by the Chinese National Space Administration (CNSA), the French Space Agency (CNES), and the Chinese Academy of Sciences (CAS). We gratefully acknowledge the unwavering support of NSSC, IAMCAS, XIOPM, NAOC, IHEP, CNES, CEA, and CNRS. This work is supported by the Strategic Priority Research Program of the Chinese Academy of Sciences(Grant No.XDB0550401)，and by the National Natural Science Foundation of China (grant Nos. 12494571 and 12494570, 12494573, 12133003). The authors are thankful for support from the National Key R\&D Program of China (grant Nos. 2024YFA161170* and 2024YFA1611700). 
\end{acknowledgements}

\bibliography{bibtex}{}

\label{lastpage}

\end{CJK}
\end{document}